%% file: hypercube.tex
\documentclass[twocolumn,showpacs,aps,floatfix]{revtex4}

\usepackage[latin1]{inputenc}
\usepackage{amsfonts}
\usepackage{amsmath}
\usepackage{amssymb}
\usepackage[american]{babel}
\usepackage{graphicx}
\usepackage{subfigure}
\usepackage{multirow}

\input{mc1}

\input{mc2}
\begin{document}

% front matter
\title{Mixing Times in Quantum Walks on the Hypercube}
\author{F.L.~Marquezino} \email{franklin@lncc.br}
\author{R.~Portugal} \email{portugal@lncc.br}

\affiliation{Laborat\'{o}rio Nacional de
Computa\c{c}\~{a}o Cient\'{\i}fica - LNCC\\Av.
Get\'{u}lio Vargas 333, Petr\'{o}polis, RJ, 25651-075, Brazil}

\author{G. Abal} \email{abal@fing.edu.uy}
\author{R. Donangelo} \email{donangel@if.ufrj.br}
\altaffiliation[Permanent address: ]{Instituto de F\'{\i}sica, Universidade Federal do Rio de Janeiro, C.P. 68528, Rio de Janeiro 21941-972, Brazil}
\affiliation{Instituto de F\'{\i}sica, Universidad de la Rep\'ublica\\
C.C. 30, C.P. 11300, Montevideo, Uruguay}

\date{\today}

\begin{abstract}
The mixing time of a discrete-time quantum walk on the hypercube is considered. The mean probability distribution of a Markov chain on a hypercube is known to mix to a uniform distribution in time $O(n\log n)$. We show that the mean probability distribution of a discrete-time quantum walk on a hypercube mixes to a (generally non-uniform)  distribution $\pi(x)$ in time $O(n)$ and the stationary distribution is determined by the initial state of the walk. An explicit expression for $\pi(x)$ is derived for the particular case of a symmetric walk. These results are consistent with those obtained previously for a continuous-time quantum walk. The effect of decoherence due to randomly breaking links between connected sites in the hypercube is also considered. We find that the probability distribution mixes to the uniform distribution as expected. However, the mixing time has a minimum at a critical decoherence rate $p\approx 0.1$. A similar effect was previously reported for the QW on the $N$-cycle with decoherence from repeated measurements of position. A  controlled amount of decoherence helps to obtain---and preserve---a uniform distribution over the $2^n$ sites of the hypercube in the shortest possible time.
\end{abstract}
\maketitle

\section{Introduction}

Quantum walks (QW) are quantum versions of Markov chains in which coherent superpositions play a key role~\cite{Kempe03}. Both the continuous-time (CTQW)  and the discrete-time (DTQW) versions of QWs~\cite{FG98, Nayak} have been used as a basis for quantum algorithms that outperform their classical counterparts~\cite{Shenvi,Amb03,Amb05,CG04,Farhi07}. Some of them are based on a QW on the hypercube, which has received considerable attention after it was shown that the mean time required to traverse the hypercube between two opposite vertexes (hitting time) is exponentially faster than its classical counterpart~\cite{Kempe02,KB05}. In this work we consider the mixing properties of a DTQW in the hypercube, both in the coherent case and in the presence of decoherence from randomly broken links. 

The probability distribution of a classical Markov chain in a hypercube mixes to the uniform distribution in time $O(n\log n)$, where $n$ is the dimension of the hypercube. In the quantum case, the probability distribution results from a unitary process and so it cannot converge to a stationary distribution. Aharonov \textit{et al} have  shown that the time-averaged distribution, defined in the discrete-time case by $\bar P(x,T)\equiv \frac{1}{T} \sum_{t=0}^{T-1} P(x,t)$,  does converge to a limiting distribution \cite{Aharonov}. One may sample from this distribution by selecting $t$ uniformly in $[0,T-1]$, running the QW for $t$ steps and then measuring the position of the walker. The time it takes for the average distribution to be $\eps$-close to this limiting distribution is known as the mixing time $M_\eps$ (we shall later provide formal definitions). Aharonov \textit{et al} give an upper bound for the mixing time of QWs on generic graphs. For a hypercube driven by a Grover coin, this bound is $O(\frac{n^{3/2}}{\eps})$. Our numerical simulations indicate a mixing time of $O(n/\eps)$, which is better than the classical mixing time. Moore and Russell have considered mixing times \textit{with respect to a uniform distribution} in both the DTQW and CTQW on hypercubes \cite{Moore}. In the case of the CTQW they found that the average limiting distribution is not uniform and that one can always find $\eps >0$ such that no mixing time to the uniform distribution exists. Since the CTQW can be obtained from a DTQW by a suitable limiting process \cite{Strauch}, we could expect the limiting distribution in the DTQW to be also non-uniform.

In this work, we show that the asymptotic distribution of a DTQW on the hypercube \textit{is not uniform}, obtain an explicit expression for it and characterize the average mixing time to this distribution. The instantaneous mixing time is a useful notion that captures the first instant in which the position distribution is $\eps-$close to a reference distribution. It has been shown that, for both kinds of QW on a hypercube, the instantaneous mixing time to the uniform distribution depends linearly on the dimension~$n$ \cite{Moore}. This represents an improvement over the corresponding classical walk, from $O(n \log n)$ to $O(n)$. We confirm this result for the DTQW and show that $\eps>0$ can be found such that an instantaneous mixing time to the stationary (non-uniform) distribution does not exist.

All these properties are affected by decoherence. The impact of decoherence on QWs has been investigated mostly in one or two-dimensional systems using repeated measurements \cite{Brun, KT02} or topological noise from broken links \cite{deco,canal-x,Amanda}. Kendon and Tregenna have presented early numerical results showing the impact of repeated measurements on the hitting time (starting at node $x$, it is the mean time required to reach the opposite node, $\bar x$) of a DTQW in a hypercube \cite{KT02}. In the presence of decoherence the natural expectation is that the position distribution of a QW in an $n$-hypercube mixes to a uniform distribution essentially as in the classical case.  Recently, the effect of repeated measurements on a QW on the hypercube has been considered analytically by Alagic and Russell using superoperator techniques for the case of the CTQW \cite{Alagic}. They found that for weak decoherence rates, both the hitting time and the mixing time (to the uniform distribution) remain linear in $n$ but if the rate of measurements is beyond a model-dependent threshold, the classical result is re-obtained, \textit{i.e.} the CTQW distribution mixes to uniform in time $O(n\log n)$. We consider here the effect on the probability distribution of the DTQW of randomly breaking links in the network. This mechanism  involves no measurements and is an example of unitary noise \cite{Biham}. 

Decoherence is usually regarded as an obstacle to quantum computing as it destroys the entanglement and superpositions required by a quantum processor. However, it has been argued that a controlled amount of decoherence might be useful to obtain particular probability distributions. For instance, weak decoherence may be used to generate a quasi-uniform distribution in the QW on a line \cite{KT02}. Maloyer and Tregenna have shown that the mixing time of a DTQW on an $N$-cycle may be reduced by allowing for some decoherence from repeated measurements, provided that those measurements affect the position of the QW \cite{Maloyer}. We find a similar effect in the case of the hypercube with decoherence from broken links. There is a critical decoherence rate for which the mixing time has a minimum. This result provides a new example in which decoherence, from topological randomness and in the absence of measurements, enhances a useful quantum feature, namely fast mixing times. 

This paper is organized as follows. In Section~\ref{sec:coherent} we consider the coherent QW on the hypercube and derive its asymptotic distribution which is non-uniform. In Section~\ref{sec:deco}, we give the unitary evolution rule in the presence of broken links and present numerical results for the mixing time as a function of the decoherence rate. In Section \ref{sec:conclusion} we summarize our main results and present our conclusions. 

\section{Coherent QW on the hypercube}
\label{sec:coherent}

A coined quantum walk in a $n$-dimensional hypercube has a Hilbert space ${\cal H}_C\otimes {\cal H}_P$, where 
${\cal H}_C$ is the $n$-dimensional coin subspace and ${\cal H}_P$ the $2^n$-dimensional position subspace. A basis for ${\cal H}_C$ is the set $\{\ket{j}\}$ for  $0\leq j\leq n-1$ and ${\cal H}_P$ is spanned by the set $\{\ket{x}\}$ with, $0 \leq x\leq 2^n-1$. A generic state of the discrete quantum walker in the hypercube is
\begin{equation}
\ket{\Psi(t)}=\sum_{j=0}^{n-1}\sum_{x=0}^{2^n-1}\psi_{j,x}(t)\ket{j,x}, \label{eq:estgeral1d}
\end{equation}
where $\ket{j,x}=\ket{j}\otimes\ket{x}$. The evolution operator for one step of the walk is
\begin{equation}\label{evol}
U=S\circ(C\otimes I),
\end{equation}
where $C=\sum_{i,j}C_{ij}\opp{i}{j}$ is a unitary coin operation in ${\cal H}_C$, $I$ is the identity in ${\cal H}_P$ and $S$ is the shift operator defined by
\begin{equation}\label{shift-coh}
S=\sum_{j=0}^n\sum_{x=0}^{2^n-1}\opp{j,x\oplus  e_j}{j,x}.
\end{equation}
Here, $x\oplus  e_j$ is the bitwise binary sum between the $n$-component binary vectors, $x=(x_{n-1}, \ldots x_1,x_0)$ and $e_j$, a null vector except for a single $1$ entry in the $j^{th}$ component. We will use both representations (decimal or binary) as needed. They may easily be recognized by the context in which they are placed.

Applying the evolution operator on state (\ref{eq:estgeral1d}) we obtain the map for the components of the wavevector,
\begin{equation}\label{eq:evolgeral1d}
\psi_{i,x}(t+1)=\sum_{j=0}^{n-1}C_{i j}\,\psi_{j,x\oplus  e_i}(t).
\end{equation}
The analysis of the problem is simplified in Fourier space. Since the hypercube is a Cayley graph of $\Z_2^n$, the most suitable transform is the Fourier transform on $\Z_2^n$, spanned by the $2^n$ kets 
$$
\ket{k}\equiv\frac{1}{\sqrt{2^n}}\sum_{x=0}^{2^n-1} (-1)^{k\cdot x}\ket{x},\qquad\qquad k\in[0,2^n-1],
$$
with bitwise product $\protect{k\cdot x\equiv\sum_{j=0}^{n-1} x_jk_j}$. The transformed amplitudes are $\protect{\tilde{\psi}_{i,k}=\frac{1}{\sqrt{2^n}}\sum_{x=0}^{2^n-1}
(-1)^{k\cdot  x}\psi_{i,x}}$.
The evolution operator is diagonal in $k$-space and acts non-trivially in the coin subspace. The operator $U_{k}$, which acts on $\ket{\Psi_{k}}\equiv\scalar{k}{\Psi}=\sum_{i=0}^{n-1}\tilde{\psi}_{i,k}\ket{i}$, has matrix elements given by $U_{k}(i,j)=(-1)^{k_i} C_{i j}$.

From this point on, let us particularize to the $n$-dimensional Grover coin, $C(i,j)\equiv 2/n-\delta_{i j}$, which obeys the permutation symmetry of the hypercube and is, among the operators of this kind, the one farthest from the identity \cite{Moore}. We shall now describe the eigenproblem of $U_k$. Its eigenvalues depend only on $n$ and on the Hamming weight of $k$, defined as $|k|\equiv\sum_{j=0}^{n-1} k_j$. However, its eigenvectors, $\ket{\nu_i(k)}$ for $i=1\ldots n$, depend on $k$.

For Hamming weights $|k|=0$ (or $|k|=n$), a set of $n-1$ degenerate eigenvectors with eigenvalue $\lambda=1$ (or $-1$) is $\ket{\nu_i(0)}=({\ket{0}-\ket{i}})/{\sqrt{2}}$, for $i\in[1,n-1]$. The remaining eigenvector, with eigenvalue $\lambda=-1$ (or $1$) is the uniform superposition  $\ket{\nu_n(0)}=\frac{1}{\sqrt{n}}\sum_{j=0}^{n-1}\ket{j}$.

In the case where the Hamming weight takes values ${0<|k|<n}$, there are $n-|k|-1$ degenerate eigenvectors with eigenvalue $-1$, given by $\ket{\nu_i(k)}=({\ket{0}-\ket{i}})/{\sqrt{2}}$ with $i\in[1,n-|k|-1]$ and $|k|-1$ degenerate eigenvectors with eigenvalue $1$, given by $\ket{\nu_i(k)}=(\ket{n-|k|}-\ket{i+1})/\sqrt{2}$ for $i\in[n-|k|,n-2]$. The two remaining eigenvalues turn out to be the most relevant. They can be expressed as 
$\protect{e^{\pm i\omega_k}}$, with $\omega_k$ defined by 
\begin{equation}\label{omega-k}
\cos\omega_k\equiv 1-\frac{2|k|}{n}.
\end{equation}
The corresponding conjugate eigenvectors are $\ket{\nu_n(k)}=\sum_{j=0}^{n-1}\alpha_j(k)\ket{j}$ and $\ket{\nu_{n-1}(k)}=\sum_{j=0}^{n-1}\alpha_j^*(k)\ket{j}$ with components $\alpha_j(k)$ given by
\begin{equation}\label{eigenvec}
\alpha_j(k)=\frac{1}{\sqrt{2}}\left(\frac{k_j}{\sqrt{|k|}}-i\frac{1-k_j}{\sqrt{n-|k|}}\right).
\end{equation}
In the last equation $i=\sqrt{-1}$. This set of normalized eigenvectors, summarized in Table~1, forms a non-orthogonal basis.

\begin{table*}
\caption{Eigenvalues and eigenstates of $U_{k}$. The quantities $\omega_k$ and $\alpha_j(k)$ are defined in Eqs.~\eqref{omega-k} and \eqref{eigenvec}, respectively. }
\begin{ruledtabular}
% use packages: array
\begin{tabular}{ccccc}
Hamming weight & Eigenvalue & Eigenstate~$\ket{\nu_i(k)}$ & index~$i$ & multiplicity \\\hline
\multirow{2}{*}{$|k|=0$}  & $-1$ & $(\ket{0}-\ket{i})/\sqrt{2}$ & $i\in[1,n-1]$ & $n-1$ \\
                          & $1$ & $\sum_{j=0}^{n-1} \ket{j}/\sqrt{n}$ & $n$ & $1$ \\\hline
\multirow{4}{*}{ $1\le|k|\le n-1$} & $-1$ & $(\ket{0}-\ket{i})/\sqrt{2}$ & $i\in[1,n-|k|-1]$ & $n-|k|-1$\\
              & $1$ & $(\ket{n-|k|}-\ket{i+1})/\sqrt{2}$ & $i\in[n-|k|,n-2]$ & $|k|-1$\\
              & $e^{i\omega_k}$ & $\sum_{j=0}^{n-1}\alpha_j(k) \ket{j}$ & $n$ & $1$\\
              & $e^{-i\omega_k}$ & $\sum_{j=0}^{n-1}\alpha_j^*(k) \ket{j}$ & $n-1$ & $1$\\\hline
\multirow{2}{*}{$|k|=n$} & $1$ & $(\ket{0}-\ket{i})/\sqrt{2}$ & $i\in[1,n-1]$ & $n-1$ \\
              & $-1$ & $\sum_{j=0}^{n-1} \ket{j}/\sqrt{n}$ & $n$ & $1$
\end{tabular}
\end{ruledtabular}
\end{table*}

\subsection{Limiting Distribution}
\label{ssec:limit-dist}

Let $P(x,t)$ be the probability to find the walker on the vertex $x$ of the hypercube at time $t$. As mentioned in the introduction, this probability depends on the initial condition and, as is typical of unitary evolutions, it does not converge to a stationary distribution. However, the time-averaged distribution $\protect{\bar P (x,T)\equiv\frac{1}{T}\sum_{t=0}^{T-1} P(x,t)}$, always converges as $T\to\infty$ \cite{Aharonov}. We define
\begin{equation}
\pi(x)\equiv\lim_{T\to\infty} \bar P(x,T)
\label{eq:Pbar}
\end{equation}%
as the stationary distribution, and consider the problem of determining it for a QW in a hypercube of dimension $n$. 
The initial statevector is localized at vertex $x=0$ and uniformly distributed in the coin subspace,
\begin{equation}\label{eq:ic}
\ket{\Psi(0)}=\frac{1}{\sqrt n}\sum_{j=1}^{n}\ket{j}\otimes \ket{x= 0}.
\end{equation}
This choice respects the permutation symmetry of the hypercube. 
The initial state is expressed in terms of the eigenvectors of $U_k$ as
$$
\ket{\Psi(0)}=\frac{1}{\sqrt{n 2^n}}\sum_{k=0}^{2^n-1}\sum_{i=1}^n a_i(k)\,\ket{\nu_i(k)}\otimes\ket{k},
$$
where the coefficients are sums of components of the eigenvectors, 
$a_i(k)\equiv \frac{1}{\sqrt{n 2^n}}\sum_{j=0}^{n-1}\scalar{\nu_i(k)}{j}$. These sums are zero unless  $i=n-1$ or $i=n$, so only these two eigenstates contribute,
$$
\ket{\Psi(0)}=\sum_{k=0}^{2^n-1}
\left(a_{n-1}(k)\ket{\nu_{n-1}(k)}+a_n(k)\ket{\nu_{n}(k)}\right)\otimes\ket{k}.
$$
The relevant coefficients are \\$\protect{a_n(k)=(\sqrt{|k|}+i\sqrt{n-|k|})/\sqrt{n\,2^{n+1}}}$ and \\ $\protect{a_{n-1}(k)=a_n^*(k)}$ when $\protect{|k|\in[1,n-1]}$. If $\protect{|k|=0}$ or $\protect{|k|=n}$, then $a_n =1/\sqrt{2^{n+1}}$ and $a_{n-1}=0$. The state of the walker at time $t$ is
\begin{multline}
\ket{\Psi(t)}=\sum_{k=0}^{2^n-1}
\left(a_{n-1}(k)\;e^{-i\omega_k t}\ket{\nu_{n-1}(k)}+\right. \\
      \left. a_n(k)\;e^{i\omega_k t}\ket{\nu_n(k)}\right)
\otimes\ket{k}.
\end{multline} %

The probability of finding the walker at time $t$ at vertex $x$ is 
$\protect{P(x,t)=\sum_{j=0}^{n-1}\left\vert\scalar{j, x}{\Psi(t)}\right\vert^2}$. This quantity can be evaluated in  the $k$-representation,
\begin{eqnarray*}
P(x,t)&=&\frac{1}{2^n}\sum_{k,k'=0}^{2^n-1}(-1)^{(k\oplus k')\cdot x}\sum_{j=0}^{n-1}\scalar{j,k}{\Psi(t)}\scalar{\Psi(t)}{k',j}\\
&=&\frac{1}{2^n}\sum_{k,k'=0}^{2^n-1}(-1)^{(k\oplus k')\cdot x}\times\\
&& \left\{a_{n-1}(k)a_{n-1}^*(k')\scalar{\nu_n(k)}{\nu_n(k')}
e^{-i(\omega_k-\omega_{k'})t}+ \right.\\
&& a_{n}(k)a_{n}^*(k')\scalar{\nu_{n-1}(k)}{\nu_{n-1}(k')}
e^{i(\omega_k-\omega_{k'})t}+\\
&& a_{n}(k)a_{n-1}^*(k')\scalar{\nu_{n-1}(k)}{\nu_{n}(k')}
e^{i(\omega_k+\omega_{k'})t}+\\
&&\left. a_{n-1}(k)a_{n}^*(k')\scalar{\nu_{n}(k)}{\nu_{n-1}(k')}
e^{-i(\omega_k+\omega_{k'})t}\right\}.
\end{eqnarray*} %
Our goal is to calculate the asymptotic probability distribution $\pi(x)$ defined by Eq.~(\ref{eq:Pbar}). The first two terms contribute only if $|k|=|k'|$, since 
$\lim_{T \to \infty}\frac{1}{T} \sum_{t=0}^{T-1} e^{\pm i(\omega_k-\omega_{k'})t}=\delta_{|k|,|k'|}$ and there is no contribution from the last two terms because   $\lim_{T\to\infty}\sum_{t=0}^{T-1}e^{\pm i(\omega_k+\omega_{k'})t}=0$. Thus, the asymptotic distribution can be evaluated from
\begin{multline}
\pi(x)=\frac{1}{2^n}\sum_{k,k'=0}^{2^n-1}\delta_{|k|,|k'|}(-1)^{(k\oplus k')\cdot x}\times\\
\left\{|a_{n-1}(k)|^2\scalar{\nu_n(k)}
{\nu_n(k')}+\right.\\
\left.|a_{n}(k)|^2\scalar{\nu_{n-1}(k)}{\nu_{n-1}(k')}\right\}.
\end{multline}

The overlap between the non-trivial eigenvectors is 
\begin{align}
\scalar{\nu_n(k)}{\nu_n(k')}&=\scalar{\nu_{n-1}(k)}{\nu_{n-1}(k')}\nonumber\\
&=\frac{n\,(k\cdot k')+|k| (n-2|k|)}{2|k|(n-|k|)}.
\end{align}
After using $|a_{n}(k)|^2=|a_{n-1}(k)|^2=1/2^{n+1}$, we obtain the stationary distribution
\begin{multline}\label{eq:pi}
\pi({x})=\frac{2}{2^{2n}}  + \frac{1}{2^{2n}}\sum_
{\begin{subarray}{c}
{k,k'=0}\\
(|k|=|k'|\neq 0,n)
\end{subarray}}^{2^n-1}  
(-1)^{(k\oplus k')\cdot x}\times\\
\left[\frac{n\,(k\cdot k')+|k| (n-2|k|)}{2|k|(n-|k|)}\right].
\end{multline}

\begin{figure*}
     %\setcaptionmargin{.5in}
\centering
\includegraphics[width=0.7\textwidth]{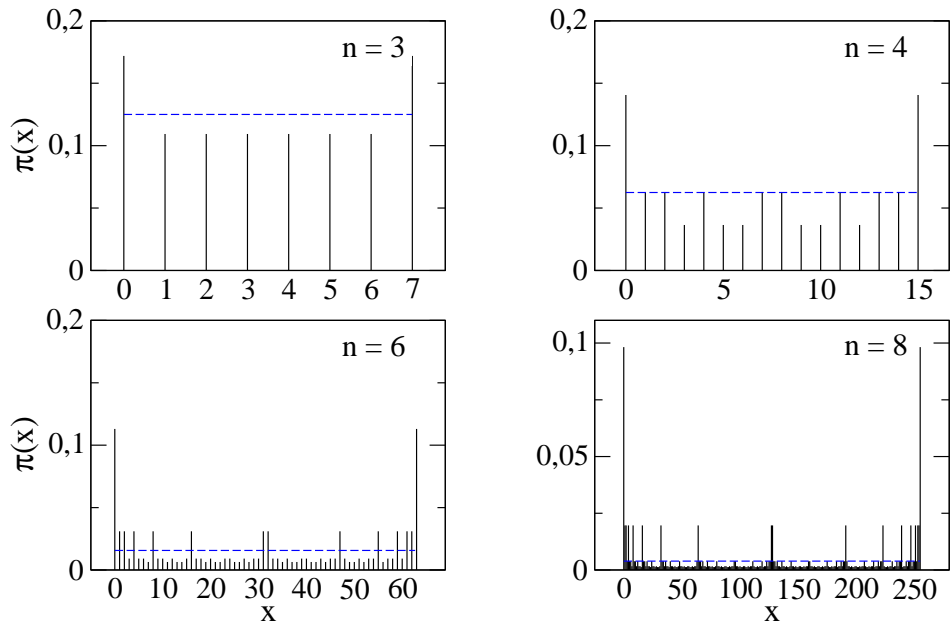} 
\caption{Limiting distributions for QW in hypercubes with $n=3,4,6,8$ obtained from Eq.~(\ref{eq:pisum}) with the initial condition~(\ref{eq:ic}). As a reference we show the uniform distribution as a dashed horizontal line. }
\label{fig01}
\end{figure*}

The above expression is not efficient to calculate $\pi({x})$. It is possible to simplify it by noting that $\pi({x})$ depends only on $|x|=\sum_{j=0}^{n-1} x_j$. After some algebra we eventually obtain
\begin{multline}\label{eq:pisum}
\pi({x}) = \frac{2}{{2}^{2\,n}} + \frac{1}{{2}^{2 n}}\sum_{i=1}^{n-1} \sum _{j=0}^{i}  \sum _{l=0}^{|x|} \sum _{m=0}^{|x|}
\sum _{{\it p}=0}^{m} (-1)^{l}\times\\
{|x|-m \choose l-m+{\it p}}{m\choose {\it p}}{n-|x|-i+m\choose i-j-l+m-{\it p}}\times\\%
{i-m\choose j-{\it p}}{n-|x|\choose i-m}{|x|\choose m}%
  \,\,\frac{i \left( n-2\,i \right) + nj }{2i \left( n-i \right) },%
\end{multline}
where the combinatorial coefficients are ${n \choose m}=\frac{n!}{(n-m)!\, m!}$ for $n\ge m\ge 0$ and ${n \choose m}=0$ otherwise. This expression is equivalent to Eq. (\ref{eq:pi}) and, for some values of $x$, it yields simple results, such as
\begin{equation}\label{eq:pi0}
\pi \left( 0 \right) = \frac{1}{4^n}+{\frac { \Gamma \left(
n+\frac{1}{2} \right) }{2 \sqrt {\pi } n \Gamma(n)} }.
\end{equation}
It should also be noted that the identity $\protect{\pi(x)=\pi(2^n-1-x)}$ also helps in the evaluation of $\pi(x)$. 
Clearly, the asymptotic distribution for $\bar P(x,t)$ is not uniform for the initial condition of Eq.\eqref{eq:ic}.
Note, for example, that for $n$ sufficiently large, Eq.~(\ref{eq:pi0}) leads to $\pi(0)\approx 1/\sqrt{2\pi(2n+1)}\gg 2^{-n}$. We have performed numerical implementations which confirm the prediction of Eq.~(\ref{eq:pisum}). This distribution is shown for hypercubes of several dimensions in Fig.~\ref{fig01}. The maximum of the distribution occurs at the initial site $x_0=0$ and at $\bar x_0=2^n-1$. Notice that $\pi(x)$ takes only $1+\lfloor n/2\rfloor$ different values which correspond to sites with Hamming distances $0,1,2\ldots \lfloor n/2\rfloor$ either to the initial site $x_0$ or to its opposite site, $\bar x_0$. 

\begin{figure}
\centering
\vspace{.1cm}
\includegraphics[width=.87\columnwidth]{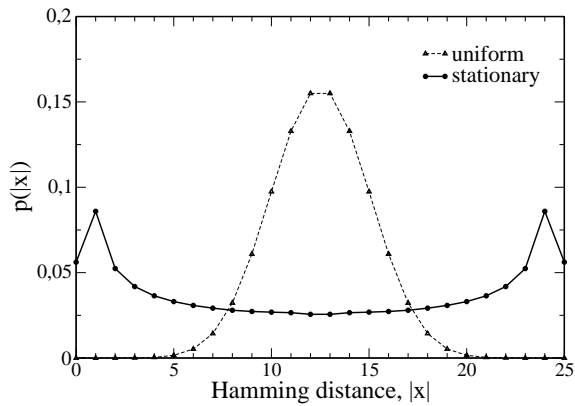} 
\caption{Asymptotic probability of finding the walker with a Hamming distance $|x|$ to the initial site, given by Eq.~(\ref{pixh}), for $n=25$. The binomial distribution $\frac{1}{2^n} {n \choose |x|}$, which corresponds to a uniform position distribution of the walker, is shown for comparison. }
\label{fig:pixh}
\end{figure}

Further insight may be obtained if one considers the probability $p(|x|)$ of finding the walker at long times at a Hamming distance $|x|$ from the initial site. There are $n\choose |x|$ sites with Hamming weight $|x|$ and this probability is simply 
\begin{equation}\label{pixh}
p(|x|)= {n\choose |x|}\pi(x).
\end{equation} 
The stationary distribution $\pi(x)$ depends only on $|x|$ and may be evaluated efficiently from Eq.~(\ref{eq:pisum}). The probability $p(|x|)$ takes at most $ 1+\lfloor n/2\rfloor$ different values, as $\pi(x)$ does. In Fig.~\ref{fig:pixh} $p(|x|)$ is compared with the corresponding probability for a uniform distribution over the hypercube, in which case the most likely situation would be to find the walker with a Hamming distance $\sim n/2$ from the starting site. Instead, we see that the most probable situation is to find it near the marked sites $x_0,\bar x_0$, with $|x|\sim 0$ or $|x|\sim n$. 

In summary, the fact that the QW starts localized at $x_0=0$ marks both $x_0$ and $\bar x_0$ as special sites and persistent interference effects give rise to a non-uniform stationary distribution over sites. The probability of finding the particle with a given Hamming distance varies much more slowly than for a uniform distribution over sites. We should remark that the results obtained for the asymptotic distribution are dependent on the initial condition. For instance, since the uniform superposition of coin \textit{and position} eigenstates, 
$\protect{\frac{1}{\sqrt{n2^n}}\sum_{j=0}^{n-1}\sum_{x=0}^{2^n-1}{\ket{j,x}}}$, is an eigenstate of $U$ with eigenvalue 1, for this particular initial condition the distribution remains uniform.

\subsection{Mixing time of a coherent evolution}
\label{ssec:coh-mix-time}

In this section we consider the mixing time for a coherent evolution. The rate at which the average probability distribution of a QW approaches its asymptotic distribution is captured by the following definition \cite{Aharonov},

\begin{definition}
 The average mixing time $M_\epsilon$ of a quantum Markov chain to a reference distribution $\pi$ is
$$M_\epsilon=\min\{T\,|\,\forall t\ge T,\norm{\bar{P}_t - \pi} \leq \epsilon \},$$ where  
$\norm{A - B}\equiv \sum_x |A(x) - B(x)|$ is the total variation distance between the two distributions.
\end{definition}
An alternative definition captures the first instant in which the walk is $\epsilon$-close to the reference distribution $\pi$, 
\begin{definition}
The instantaneous mixing time $I_\epsilon$ of a quantum Markov chain is
$$I_\epsilon=\min\{t\,|\,\norm{P_t - \pi} \leq \epsilon\}.$$
\end{definition}
Both mixing times depend on the initial condition of the QW. 

\begin{figure*}
\centering
\includegraphics[height=.9\columnwidth, angle=270]{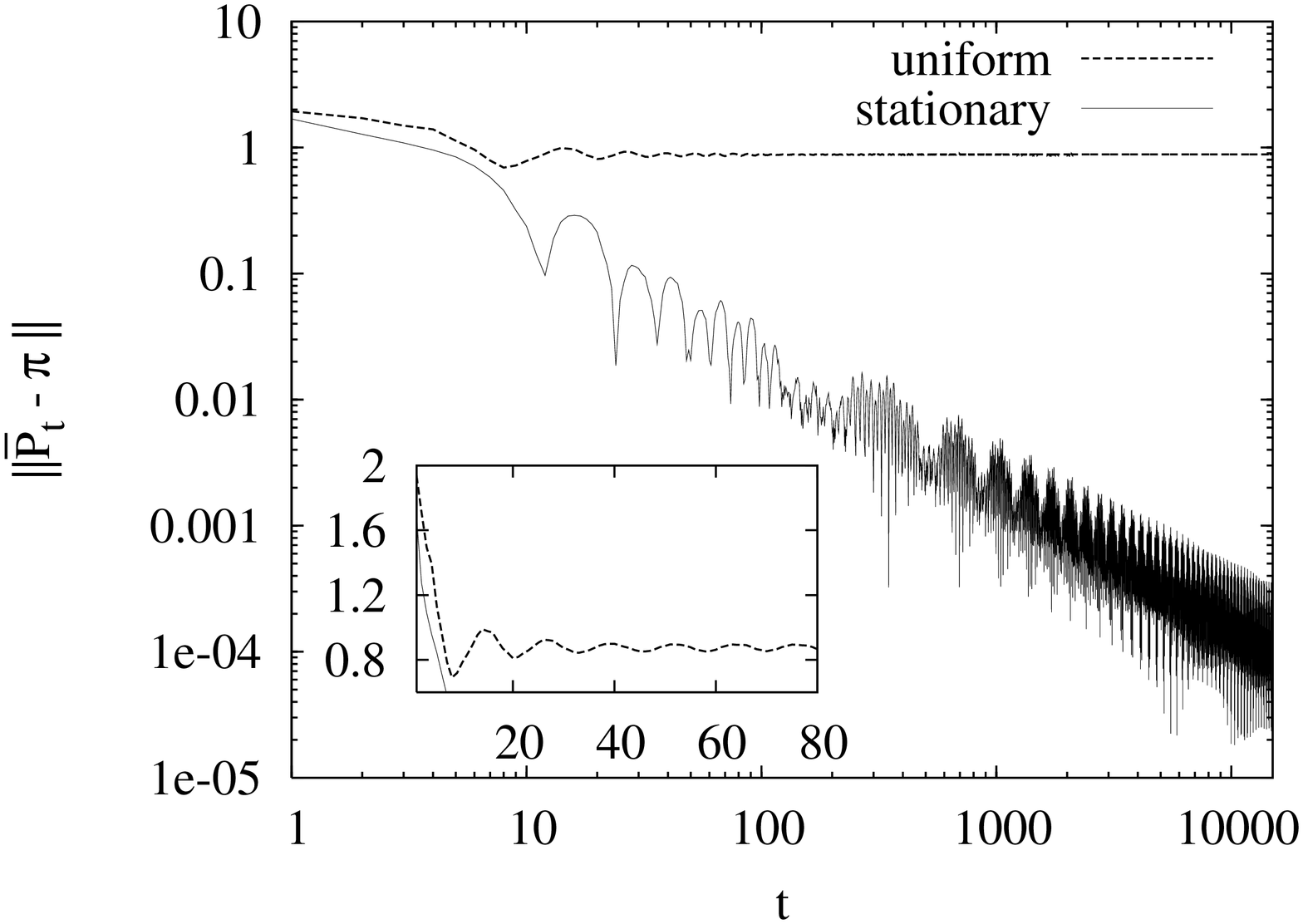}~
\includegraphics[height=.86\columnwidth, angle=270]{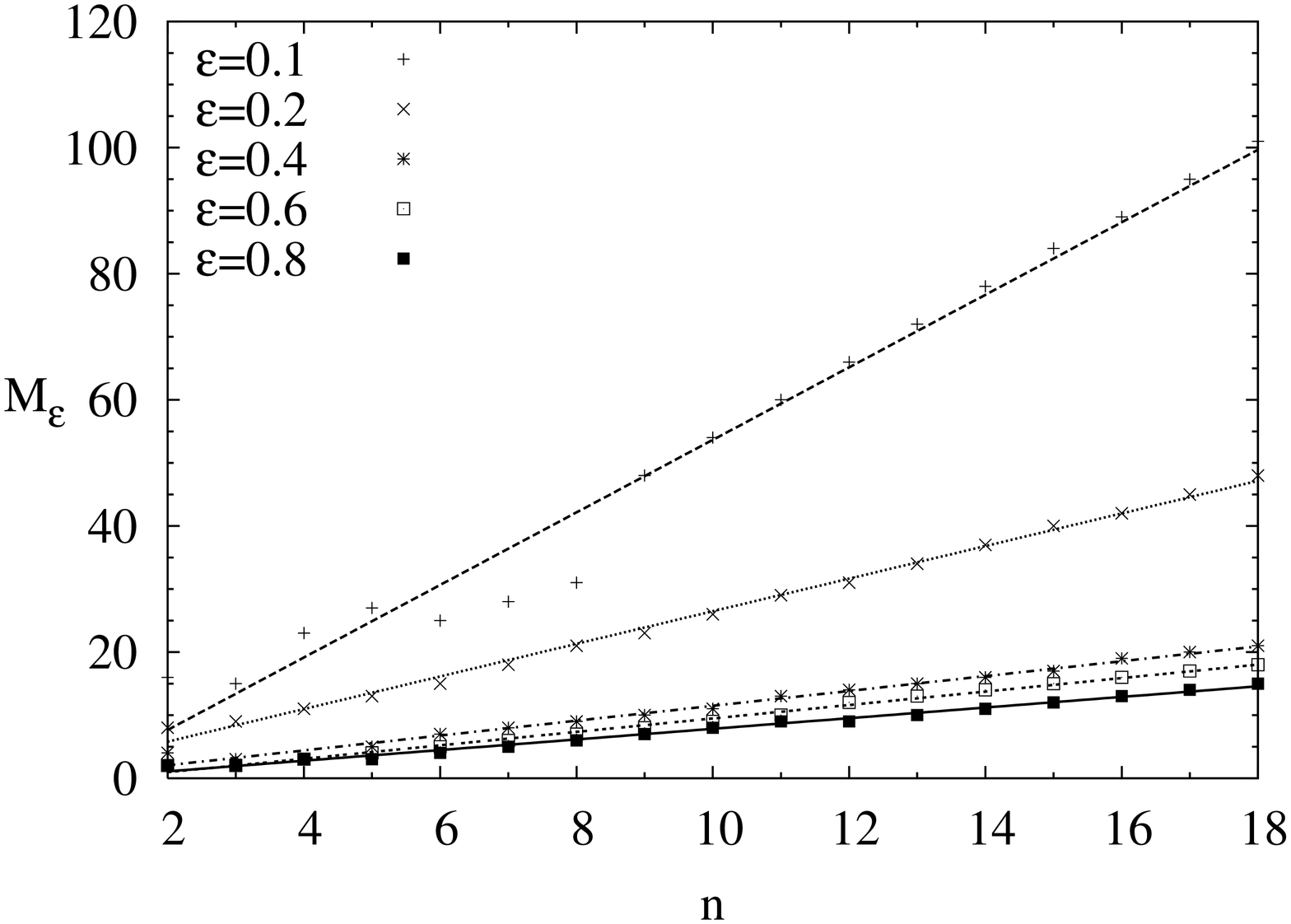}
\caption{Left: total variation distance from the average distribution at time $t$ to both the uniform and the stationary distributions, for a coherent QW moving on a hypercube of dimension $n=8$. Axes are logarithmic in the larger plot and linear in the inset. Right: Mixing time as a function of the dimension $n$ for several thresholds $\epsilon$.} 
\label{fig:TVDn8}
\end{figure*}

The left panel of Fig.~\ref{fig:TVDn8} shows the time dependence of the variation distance of the average distribution to the stationary distribution, Eq.~(\ref{eq:pi}). For long times, the variation distance decays approximately as $\sim 1/t$ while the corresponding distance to the uniform distribution remains essentially constant. 

For a quantum walk in a generic graph with arbitrary initial condition, an upper bound for the total variation distance to the asymptotic distribution $\pi(x)$ was derived by Aharonov \textit{et al.}~\cite{Aharonov}, 
$$
\|\bar P(x,T) - \pi(x)\|\leq \frac{\pi}{T\Delta}\left(1+\ln(n2^{n-1})\right).
$$
This bound is expressed in terms of the minimum separation $\Delta$ between distinct eigenvalues of $U$. For a hypercube of dimension $n$ driven by a Grover coin, this minimum separation is $\protect{\Delta=\min|e^{i\omega_k}-1|=2/\sqrt{n}}$. Thus, for large $n$, the average mixing time $M_\epsilon$ is bounded by $O\left(\frac{n^{3/2}}{\epsilon}\right)$. The right panel of Fig.~\ref{fig:TVDn8} shows the linear dependence of the average mixing time with the dimension $n$ and the threshold $\epsilon$, for the initial state given by Eq.~(\ref{eq:ic}). For sufficiently large dimension, it scales as $n/\eps$, which is consistent with this bound. 

\begin{figure*}
\centering
\includegraphics[height=.9\columnwidth, angle=270]{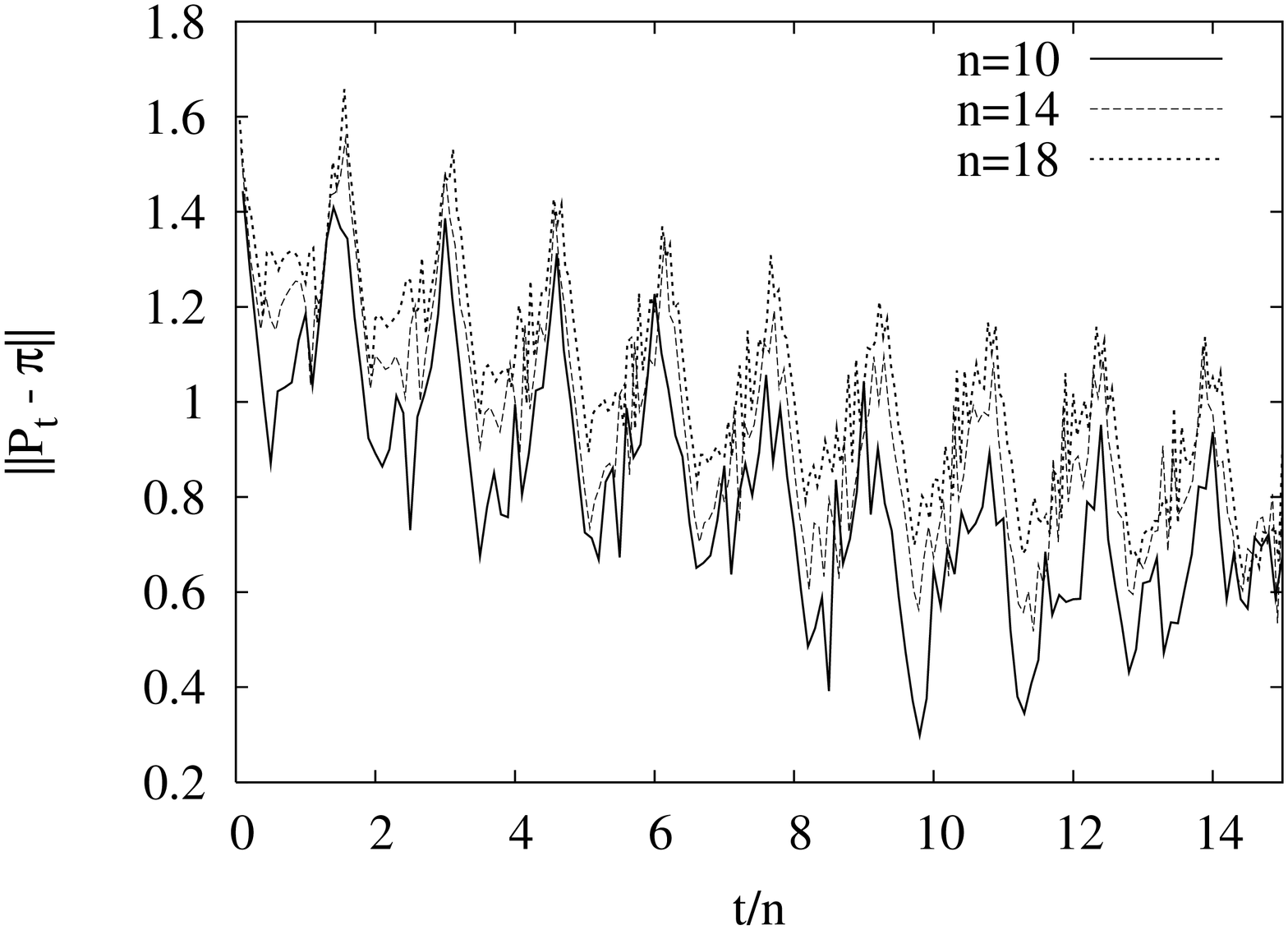}~
\includegraphics[height=.9\columnwidth, angle=270]{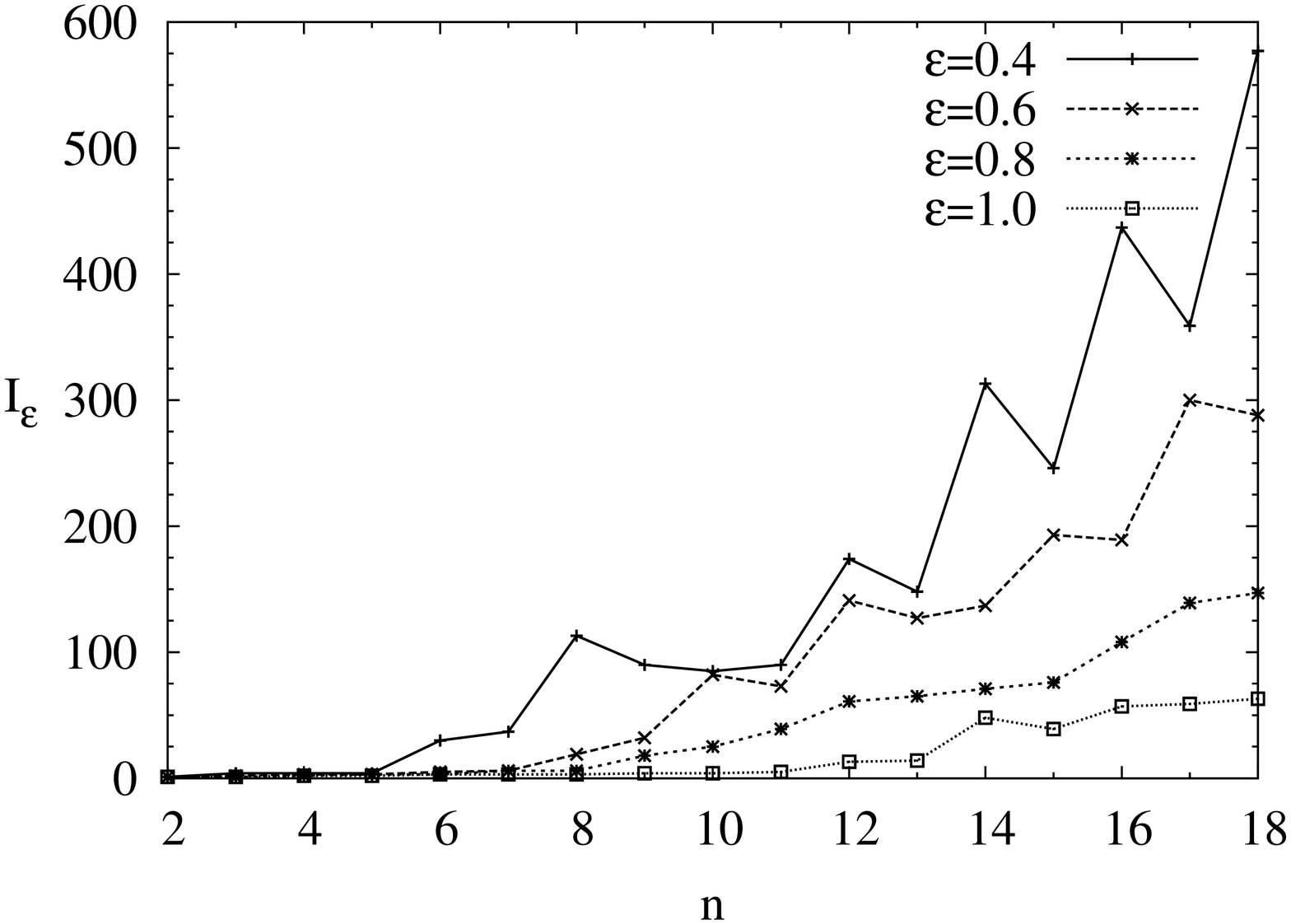}
\caption{Left panel: Total variation distance to the stationary distribution $\pi(x)$, as a function of $t/n$.  
Right panel: Instantaneous mixing time $I_\epsilon$ to the stationary distribution as a function of the dimension $n$.}
\label{fig:MIX-Sinst}
\end{figure*}

\begin{figure}
\centering
\includegraphics[height=.9\columnwidth, angle=270]{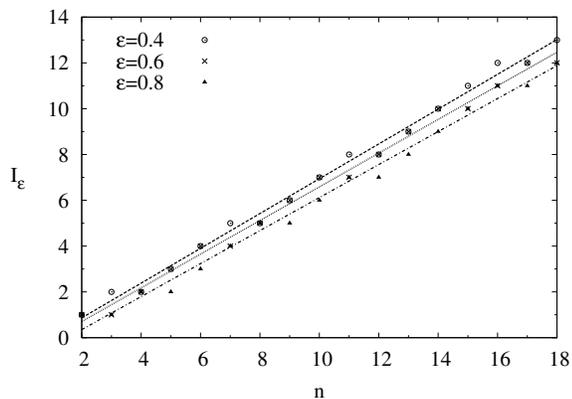}
\caption{Instantaneous mixing time to the uniform distribution as a function of the dimension $n$.} 
\label{fig:MIX-Uinst}
\end{figure}

The instantaneous mixing time, $I_\epsilon$, the first time for which the position distribution $P(x,t)$ is $\epsilon$-close to a given distribution, must be calculated considering the appropriate parity of each step of the walk. In the left panel of Fig.~\ref{fig:MIX-Sinst} we show the total variation distance of the instantaneous distribution to the stationary distribution $\pi(x)$, given by Eq.~(\ref{eq:pisum}), as a function of $t/n$, for several values of $n$.  Note that the local minima do not approach zero as $t$ increases. Hence, for $\epsilon\lesssim 0.3$ and fixed $n$, there is no $\epsilon$-instantaneous mixing time. In the right panel of Fig.~\ref{fig:MIX-Sinst} we show the instantaneous mixing time as a function of the dimension $n$ of the hypercube. There are large oscillations in $I_\eps$ for small $\eps$ and the dependence on $n$ is clearly nonlinear. 

Moore and Russell \cite{Moore} explore the instantaneous mixing time to the uniform distribution with appropriate parity. This is a useful notion that captures the first instant in which the distribution of the QW is $\epsilon$-close to uniform. In Fig.~\ref{fig:MIX-Uinst} we show this  $\epsilon$-instantaneous mixing time as a function of the dimension $n$ of the hypercube, for several thresholds. It scales linearly with $n$ with a slope close to $\pi/4$. As reported in \cite{Moore} ---and confirmed by our numerical calculations---, for $t/n=\pi /4$, the position distribution of the QW on the $n$-hypercube is close to uniform. However, the variation distance to the stationary distribution $\pi(x)$, does not have this property, as shown in the left panel of Fig.~\ref{fig:MIX-Sinst}.

\section{Decoherence and Mixing Times}
\label{sec:deco}

Let us now consider the effects of decoherence on the stationary distribution and on the average mixing time of the QW on a hypercube. As a source of decoherence we consider topological noise which randomly opens links between connected sites of the hypercube. The effects of this broken-link noise model have been previously considered for a QW on a line \cite{deco,canal-x} and a on plane \cite{Amanda}. The broken-link noise is an example of ``unitary noise'' \cite{Biham} which may affect the dynamics in a different way from that resulting from performing coin or position partial measurements with probability $p$. This type of noise is characterized by a sequence of uncorrelated unitary operations applied on an initial state, 
\begin{equation}\label{unitary-noise}
 \ket{\Psi(t)}=U_tU_{t-1}\ldots U_1\ket{\Psi(0)}
\end{equation} 
and no measurements are performed during the evolution. Each operator $U_i$, for $i=1,\ldots t$, is of the form of Eq.~(\ref{evol}), with a modified shift operator, $S^\prime$, which accounts for the current state of the network, \textit{i.e.} which links are broken at time step $i$.  

\begin{figure*}
\centering
\includegraphics[width=.8\textwidth]{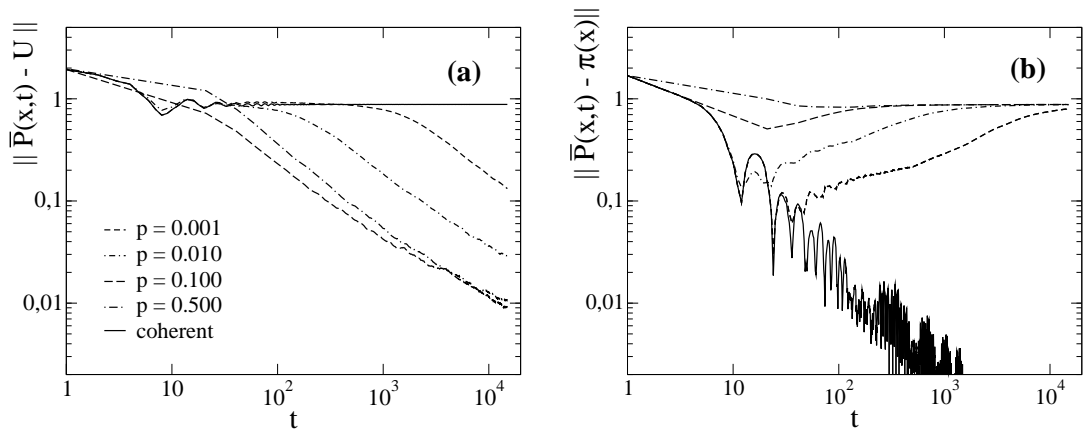}
\caption{Evolution of the total variation distance of the average distribution to (a) the uniform distribution, $U=2^{-n}$ and (b) the stationary distribution, $\pi(x)$ from Eq.~(\ref{eq:pi}). Several decoherence rates are shown, together with the coherent case $(p=0)$, for fixed dimension $n=8$.}
\label{fig:TVD-decoherent8}
\end{figure*}

The generalization of the shift operator should preserve unitarity. We take as a starting point the shift operator defined in Eq.~(\ref{evol}), 
\begin{equation}\label{shift1}
S=\sum_{j=0}^{n-1}\sum_{x=0}^{2^n-1}\opp{x\oplus  e_j}{x}\otimes\op{j},
\end{equation}
where $e_j$, a binary $n$-component vector with $|e_j|=1$, labels the $n$ spatial directions in the hypercube. The simplest way to generalize $S$ to include the possibility of broken links is to define site-dependent, $n$-component binary vectors  
\begin{align}\label{bl}
 e_j^\prime(x)&=e_j^\prime(x\oplus e_j)\\
              &\equiv
\begin{cases}
 e_j,&\mbox{if link at site $x$ in direction $j$ is closed,}\\
0,& \mbox{if link at site $x$ in direction $j$ is open.}
\end{cases}\nonumber
\end{align} 
Then, the modified shift operator is simply
\begin{equation}\label{shift-bl}
 S^\prime\equiv\sum_{j=0}^{n-1}\sum_{x=0}^{2^n-1}\opp{x\oplus e_j^\prime(x)}{x}\otimes\op{j}.
\end{equation}  
If at site $x$ the link in direction $j$ is broken, $\protect{\opp{x\oplus  e_j^\prime}{x}=\opp{x}{x\oplus  e_j}=\op{x}}$ and no probability flux is transferred across the broken link. The modified shift operator $S^\prime$ is unitary for any number of broken links. The evolution proceeds as follows. At each time step, the topology of the hypercube is defined, opening each link with probability $p$ and setting the vectors $e_j^\prime(x)$ according to Eq.(\ref{bl}). Then $S^\prime\cdot (I\otimes C)$ is applied to $\ket{\Psi(t)}$ to generate the state at time $t+1$. Note that in this model the state of the network at time $t+1$ is uncorrelated with the previous state at $t$. 

\begin{figure*}
\centering
\includegraphics[height=.9\columnwidth, angle=270]{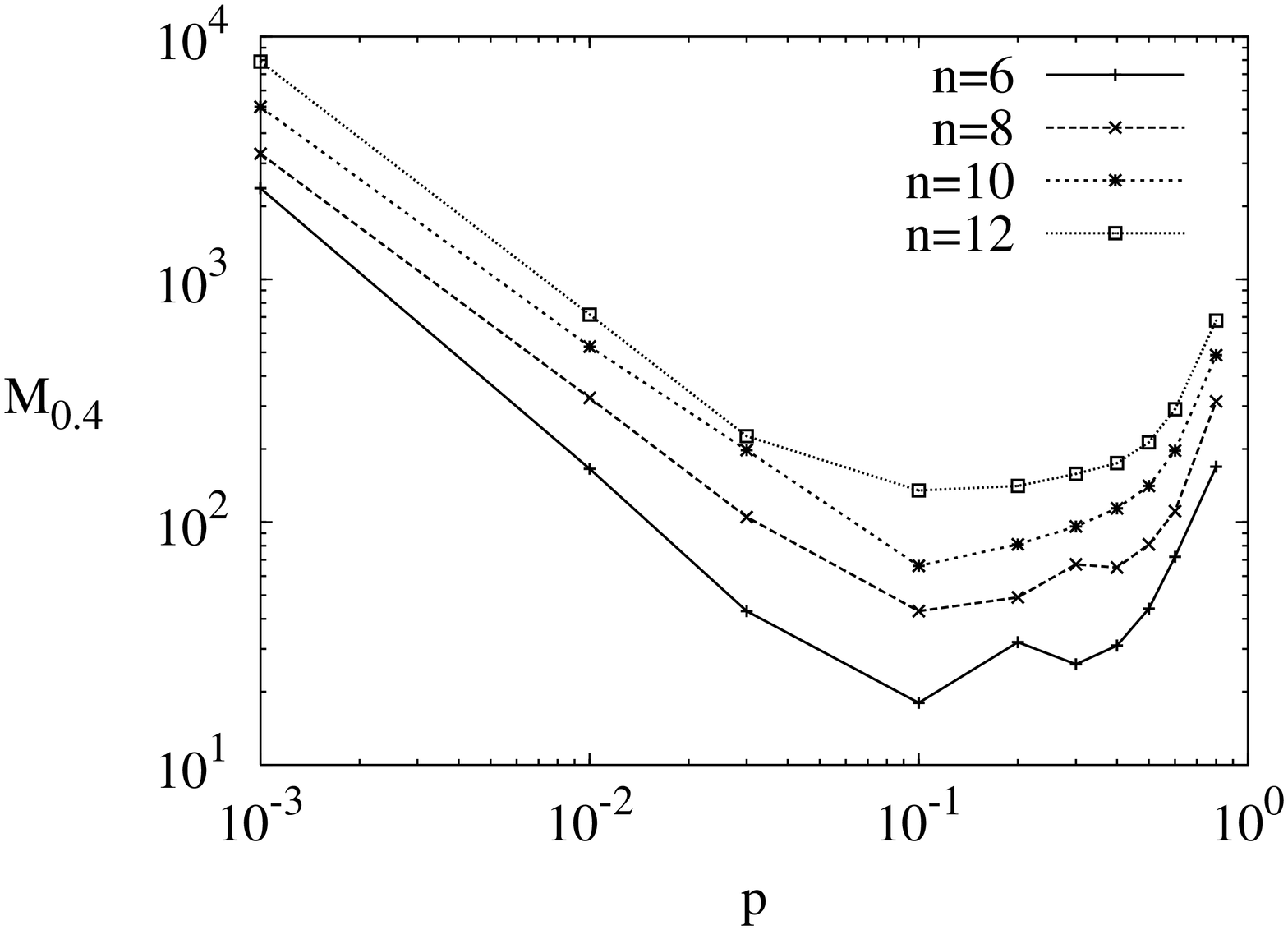}~
\includegraphics[height=.9\columnwidth, angle=270]{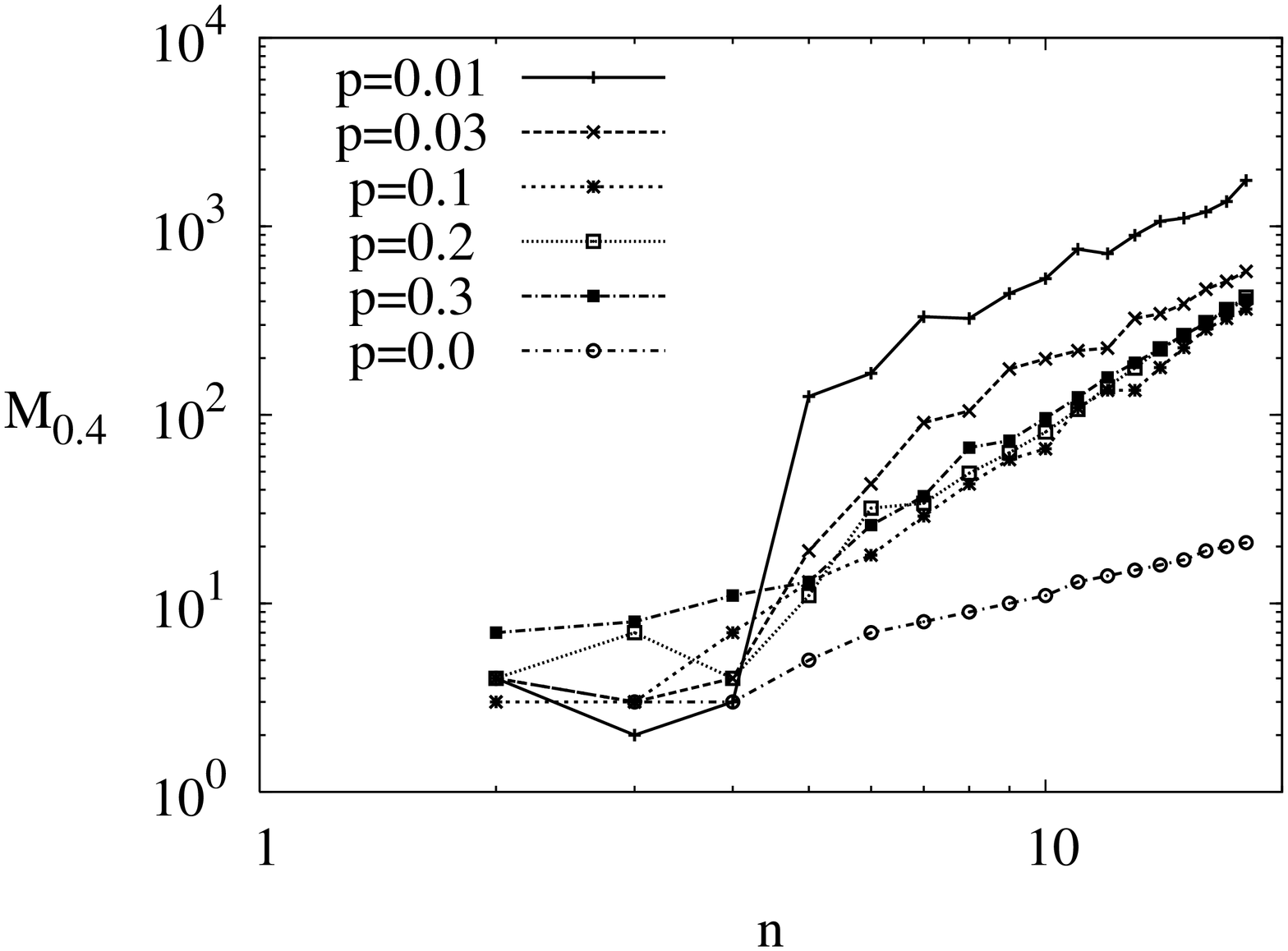}
\caption{Left panel: Average mixing time to the uniform distribution in a decoherent hypercube as a function of the probability of broken links $p$. Right panel: the same quantity as a function of dimension $n$. The average mixing time to the stationary distribution for the coherent case is also shown for comparison (curve with circles).}
\label{fig:MIX-eps}
\end{figure*}

In the presence of decoherence, the stationary distribution in the $n$-hypercube is independent of the initial conditions. In Fig.~\ref{fig:TVD-decoherent8} we show the evolution of the total variation distance to (a) the uniform distribution and (b) the stationary distribution of the coherent case, Eq.~(\ref{eq:pi}). It is clear that the introduction of even weak decoherence causes the asymptotic distribution to become uniform in a characteristic time $\sim 1/p$. After this time, the total variation distance to the uniform distribution decays according to an inverse power law which is independent of the decoherence rate $p$. For weak decoherence rates, the average probability distribution remains close to that of the coherent case for a time of order $p^{-1/2}$, as shown in panel (b) of Fig.~\ref{fig:TVD-decoherent8}. 

It is also apparent from Fig.~\ref{fig:TVD-decoherent8} that the case $p=0.1$ goes faster to the uniform distribution than the other cases shown, both for smaller and larger values of $p$. In fact,  when the decoherence rate is reduced or increased around this critical value, the convergence rate to the uniform distribution becomes slower. The fact that a  critical decoherence rate $p_c$ minimizes the mixing time has been previously reported by Kendon and Tregenna \cite{KT02} in the context of a decoherent QW in the $N-$cycle. The dependence of the mixing time of the decoherent QW on the hypercube on the decoherence rate or, equivalently, the probability $p$ of broken links is shown in 
the left panel of Fig.~\ref{fig:MIX-eps}. A minimum can be identified near $p_c\approx 0.1$, which corresponds to a decoherence rate that may provide a faster mixing time in the hypercube. The critical value appears to be independent, or, at most, weakly dependent, on the dimension of the hypercube. This conclusion is similar to the one obtained by Kendon and Tregenna~\cite{KT02} for the $N$-cycle with repeated measurements, although we have employed here a different kind of decoherence on a very different network.

In the right panel of Fig.~\ref{fig:MIX-eps} we show the dependence on the dimension of the average mixing time to the uniform distribution for different decoherence levels. The linear mixing time to the stationary distribution, Eq.~\eqref{eq:pisum}, of the coherent case is also shown (curve with circles). The decoherent mixing times increase with dimension at a rate which is approximately $n^{7/3}$, i.e. slightly faster than quadratic, so that the mixing times with broken links are larger than the coherent mixing time for all dimensions. Note that the data in Fig.~\ref{fig:MIX-eps} also confirms that the walk with $p\approx 0.1$ mixes faster than the walks with other decoherence rates. 

\section{Discussion}
\label{sec:conclusion}

In this paper the mixing properties of a discrete-time quantum walk on the $n$-dimensional hypercube has been considered in detail. The effect of decoherence from broken links --- a specific kind of unitary noise which involves no measurements --- on the mixing properties of the walk has also been investigated. 

The stationary distribution for the coherent quantum walk on the $n$-hypercube driven by a Grover coin has been found analytically, for the particular case of a symmetric initial condition. It is \emph{not} the uniform distribution.
%as reported by Moore and Russell~\cite{Moore}. 
However, the stationary distribution in this case is such that all \emph{Hamming weights} are roughly equiprobable. It has also been noted that this distribution is dependent on the initial condition.

According to our numerical simulations the mixing time $M_\eps$ on the $n$-hypercube increases as $O(n/\eps)$. We have shown this fact to be consistent with a general result of Aharonov \textit{et al.}~\cite{Aharonov} which, when particularized for the $n$-dimensional hypercube, provides an upper bound of $O(n^{3/2}/\eps)$.

The total variation distance to the uniform distribution shows a local minimum for $t=\frac{\pi}{4}n$, as reported by Moore and Russell~\cite{Moore}, but a similar behavior with respect to the stationary distribution has not been observed. We have also found that the instantaneous mixing time to the stationary distribution,  when it exists, has a nonlinear dependence on the dimension of the hypercube.
These results reconcile the discrete-time quantum walk with the continuous-time quantum walk on the hypercube. The average distributions of both walks on the $n$-hypercube fail to converge to the uniform distribution, but both are uniform or $\eps$-close to uniform at certain times ($t=\frac{\pi}{4}n$).

Decoherence, even at low rates, causes the stationary distribution to be the uniform distribution. The decay of the total variational distance takes place after a characteristic time $1/p$ and follows an inverse power law independent of $p$. For the case of broken-links unitary noise, which involves no measurements, an optimal decay rate for which the mixing time is a minimum has been found. In the mixing time as a function of dimension, the same optimal decay rate has been found to provide a faster convergence in the decoherent case. A similar effect has been reported by Kendon and Tregenna~\cite{KT02} in the $N$-cycle with decoherence from frequent partial measurements.

In future work we are interested in generalizing these results in two ways: (i) finding the stationary distributions induced by arbitrary initial coins and (ii) considering different coin operations, specially the Fourier coin.

\begin{acknowledgements}
G.A. acknowledges the warm hospitality of the LNCC, Petr\'opolis, Brazil,
where this work was initiated. G.A. and R.D. aknowledge financial
support from PEDECIBA (Uruguay), Plan de Desarrollo Tecnol\'ogico PDT
Proy. S/C/IF 54 (Uruguay) and from the Millenium Institute for Quantum
Information. R.D. also acknowledges financial support from the
Conselho Nacional de Desenvolvimento Cient\'{\i}fico e Tecnol\'ogico (CNPq),
and the Funda\c{c}\~ao de Amparo \`a Pesquisa do Estado do Rio de Janeiro (FAPERJ), Brazil.
F.L.M. acknowledges financial support from CNPq.
\end{acknowledgements}

\bibliographystyle{h-physrev} % PHYS REV adapted for arXiv preprints
\bibliography{hypercube}

\end{document}

%% file: mc1.tex
\newcommand{\mat}{\left( \begin{array}{cc}}
\newcommand{\rix}{\end{array} \right)}

\newcommand{\Z}{\mathbb{Z}}

\newcommand{\eps}{\epsilon}

\newcommand{\norm}[1]{\left\| #1 \right\|}

\newtheorem{definition}{Definition}[section]

%% file: mc2.tex
\newcommand{\ket}[1]{|#1\rangle}

\newcommand{\scalar}[2]{\langle#1|#2\rangle}

\newcommand{\op}[1]{|#1\rangle\langle#1|}
\newcommand{\opp}[2]{|#1\rangle\langle#2|}

%% file: hypercube.bbl
\begin{thebibliography}{10}

\bibitem{Kempe03}
J.~Kempe,
\newblock Contemp. Phys. {\bf 44}, 307 (2003), arXiv:quant-ph/0303081v1.

\bibitem{FG98}
E.~Farhi and S.~Gutmann,
\newblock Phys. Rev. A {\bf 58}, 915 (1998), arXiv:quant-ph/9706062v2.

\bibitem{Nayak}
A.~Nayak and A.~Vishwanath,
\newblock Quantum walk on a line, 2000, arXiv:quant-ph/0010117,
\newblock DIMACS Technical Report 2000-43.

\bibitem{Shenvi}
N.~Shenvi, J.~Kempe, and K.~B. Whaley,
\newblock Phys. Rev. A {\bf 67}, 052307 (2003), arXiv:quant-ph/0210064.

\bibitem{Amb03}
A.~Ambainis,
\newblock Quantum walk algorithm for element distinctness,
\newblock in {\em Proceedings 45th Annual IEEE Symp. on Foundations of Computer
  Science (FOCS)}, 2004, arXiv:quant-ph/0311001.

\bibitem{Amb05}
A.~Ambainis,
\newblock Quantum search algorithms, 2005, arXiv:quant-ph/0504012.

\bibitem{CG04}
A.~Childs and J.~Goldstone,
\newblock Phys. Rev. A {\bf 70}, 022314 (2004), arXiv:quant-ph/0306054.

\bibitem{Farhi07}
E.~Farhi, J.~Goldstone, and S.~Gutmann,
\newblock A quantum algorithm for the hamiltonian nand tree,
  arXiv:quant-ph/0702144v2.

\bibitem{Kempe02}
J.~Kempe,
\newblock Quantum random walks hit exponentially faster,
\newblock in {\em Proceedings of 7th International Workshop on Randomization
  and Approximation Techniques in Computer Science (RANDOM 03)}, pp. 354--369,
  2003, arXiv:quant-ph/0205083.

\bibitem{KB05}
H.~Krovi and T.~Brun,
\newblock Phys. Rev. A {\bf 73}, 032341 (2006), arXiv:quant-ph/0510136v1.

\bibitem{Aharonov}
D.~Aharonov, A.~Ambainis, J.~Kempe, and U.~Vazirani,
\newblock Quantum walks on graphs,
\newblock in {\em Proceedings of 33th ACM Symposium on Theory of Computation
  (STOC'01)}, pp. 50--59, New York, NY, 2001, ACM, arXiv:quant-ph/0012090v2.

\bibitem{Moore}
C.~Moore and A.~Russell,
\newblock Quantum walks on the hypercube,
\newblock in {\em Proceedings of 6th International Workshop on Randomization
  and Approximation Techniques (RANDOM 2002), Vol. 2483 of Lecture Notes in
  Computer Science (LNCS)}, edited by J.~D.~P. Rolim and S.~Vadhan, pp.
  164--178, Cambridge, MA, 2002, Springer-Verlag, Berlin, 2002,
  arXiv:quant-ph/0104137v1.

\bibitem{Strauch}
F.~Strauch,
\newblock Phys. Rev. A {\bf 74}, 030301 (R) (2006), arXiv:quant-ph/0606050.

\bibitem{Brun}
T.~Brun, H.~Carteret, and A.~Ambainis,
\newblock Phys. Rev. A {\bf 67}, 032304 (2003), arXiv:quant-ph/0210180.

\bibitem{KT02}
V.~Kendon and B.~Tregenna,
\newblock Phys. Rev. A {\bf 67}, 042315 (2003), arXiv:quant-ph/0209005.

\bibitem{deco}
A.~Romanelli, R.~Siri, G.~Abal, A.~Auyuanet, and R.~Donangelo,
\newblock Physica A {\bf 347}, 137 (2004), arXiv:quant-ph/0403192.

\bibitem{canal-x}
G.~Abal, R.~Donangelo, F.~Severo, and R.~Siri,
\newblock Physica A {\bf 387}, 335 (2007), arXiv:quant-ph/0708.1297.

\bibitem{Amanda}
A.~Oliveira, R.~Portugal, and R.~Donangelo,
\newblock Phys. Rev. A {\bf 74}, 012312 (2006).

\bibitem{Alagic}
G.~Alagic and A.~Russell,
\newblock Phys. Rev. A {\bf 72}, 062304 (2005), arXiv:quant-ph/0501169.

\bibitem{Biham}
D.~Shapira, O.~Biham, A.~Bracken, and M.~Hackett,
\newblock One dimensional quantum walk with unitary noise,
\newblock 2003.

\bibitem{Maloyer}
O.~Maloyer and V.~Kendon,
\newblock New J. Phys. {\bf 9}, 87 (2007), arXiv:quant-ph/0612229.

\end{thebibliography}
